\documentclass[prd,superscriptaddress,nofootinbib,notitlepage]{revtex4-1}
\usepackage{epsfig,amsmath,amssymb,slashed}
\usepackage{graphicx}
\usepackage{color}
\usepackage{siunitx}
\usepackage{color}
\usepackage{bm}
\usepackage{subfig}
\newcommand{\bra}[1]{\langle #1|}
\newcommand{\ket}[1]{|#1\rangle}

\newcommand{\di}{{\rm d}}
\newcommand{\ii}{i}

\def\wT{{\widehat T}}

\def\wpsi{{\widehat{\psi}}}
\def\wchi{{\widehat{\chi}}} 
\def\wrho{{\widehat{\rho}}}

\def\wa{\widehat a}
\def\wad{\widehat a^{\dagger}}

\def\wc{\widehat c}
\def\wcd{\widehat c^{\dagger}}

\newcommand{\Tr}{{\rm Tr}}  
  
\newcommand{\e}{{\rm e}}
\newcommand{\kk}{{\rm k}}

\newcommand{\x}{{\rm x}}

\newcommand{\be}{\begin{equation}}
\newcommand{\ee}{\end{equation}}                                                                               
\def\bea{\begin{eqnarray}}
\def\eea{\end{eqnarray}}      
\begin{document}

\title{Quantum field corrections to the equation of state of freely streaming matter 
in the Friedman-Lema\^itre-Robertson-Walker space-time} 

\author{F. Becattini \footnote[1]{Corresponding author. E-mail address: becattini@fi.infn.it}}
\affiliation{Universit\`a di Firenze and INFN Sezione di Firenze, Florence, Italy}
\author{D. Roselli}
\affiliation{Universit\`a di Firenze and INFN Sezione di Firenze, Florence, Italy}

\begin{abstract}
We calculate the energy density and pressure of a scalar field after its decoupling from a thermal 
bath in the spatially flat Friedman-Lema\^itre-Robertson-Walker space-time, within the framework of 
quantum statistical mechanics. By using the density operator determined by the condition of local thermodynamic equilibrium, we calculate the mean value of the stress-energy 
tensor of a real scalar field by subtracting the vacuum expectation value at the time of the decoupling. 
The obtained expressions of energy density and pressure involve corrections with respect to the classical 
free-streaming solution of the relativistic Boltzmann equation, which may become relevant even at long 
times.
\end{abstract}

\maketitle

\section{Introduction}
\label{sect1}

A key ingredient in cosmology is the equation of state of matter, i.e. the relation between energy 
density $\varepsilon$ and the pressure $p$. They are the only two independent scalar quantities 
appearing in the stress-energy tensor that are allowed by the symmetries of the Friedman-Lema\^itre-Robertson-Walker 
(FLRW) metric. In the comoving cosmological coordinates this reads:
\be\label{frw}
 \di s^2 = \di t^2 - a(t)^2 \left( \frac{\di r^2}{1-Kr^2} + r^2 \di \Omega^2 \right)  
\ee
where $a(t)$ is the scale factor and $K=0,\pm1$. Indeed, the isotropy and homogeneity of the 
Universe, encoded in the metric \eqref{frw}, dictates that the rank 2 symmetric stress-energy 
tensor can only be of the perfect fluid form:
\be\label{set}
 T^{\mu\nu} = (\varepsilon+p) \, u^\mu u^\nu - p g^{\mu\nu}
\ee
where $u^\mu = (1,{\bf 0})$. 

The equation of state of ordinary (or dark) matter is usually inferred by imposing thermodynamic 
equilibrium relations. Strictly speaking, thermodynamic equilibrium is possible only in stationary 
space-times, endowed with a global Killing time-like vector. Indeed, thermal field theory in 
curved stationary space-times has been the subject of several studies in literature 
\cite{dowker1,dowker2,critchley,miele}. However, in an expanding Universe described
by the FLRW metric \eqref{frw} only local thermodynamic equilibrium can be defined. Local equilibrium
applies as long as matter is coupled within the cosmological plasma; when matter, in the form of a specific 
quantum field, decouples because the expansion rate exceeds the reaction rate \cite{kolbturner}
particle spectra get frozen thereafter. The freeze-out process is usually studied by means of
the classical relativistic kinetic theory, specifically the relativistic Boltzmann equation 
\cite{kremer}. After freeze-out, the classical phase space distribution function $f(x,k)$ is the 
so-called free-streaming solution of the Boltzmann equation and the relation between stress-energy 
tensor and distribution function reads \cite{piattella}:
$$
  T^{\mu\nu} = \int \di k_1 \di k_2 \di k_3 \; \frac{k^\mu k^\nu}{k^0 \sqrt{-g}} f(x,k) 
$$
where $k$ is the four-momentum of the particle. Since the covariant components $k_\mu$ are
conserved in the cosmological free fall, the solution of the relativistic Boltzmann equation 
is simply:
$$
  f(t,x^1,x^2,x^3,k_\mu) = f_0(t_0,x^1,x^2,x^3,k_\mu)
$$
where $f_0$ is the distribution function at the time $t_0$. Thus, in the approximation of a sudden 
freeze-out, that is of an instantaneous transition from local thermodynamic equilibrium to a non 
interacting system, the energy density and pressure of, e.g., freely streaming neutral spinless 
particles read:
\be\label{cfs}
\begin{split}
 \varepsilon(t) &= \frac{1}{(2\pi)^3 a^4(t)} \int \di^3 \kk \; \sqrt{\kk^2 + m^2 a^2(t)} 
  \frac{1}{\e^{\sqrt{\kk^2+m^2}/T(t_0)}-1}  \\
 p(t) &= \frac{1}{3(2\pi)^3 a^4(t)} \int \di^3 \kk \; \frac{\kk^2}{\sqrt{\kk^2+m^2a^2(t)}}
 \frac{1}{\e^{\sqrt{\kk^2+m^2}/T(t_0)}-1} 
\end{split}
\ee
where ${\bf k}$ are the space covariant components, in the metric \eqref{frw}, of the particle 
four-momentum vector (which are conserved in the free-streaming), and $\kk^2 = k_x^2+k_y^2+k_z^2$; 
$T(t_0)$ is the temperature at the decoupling, or freeze-out in this approximation, and the scale 
factor $a(t_0)$ at the freeze-out time $t_0$ is set to be 1, which will be henceforth assumed.

In this work, we will show that a proper quantum statistical handling implies corrections to the
energy density and pressure \eqref{cfs} expressions of the free-streaming solution for a sudden
freeze-out which, in principle, can become relevant also at late times.  We will show that such 
corrections arise from the interplay of quantum statistical mechanics and the evolution of free 
fields in a curved spacetime. If such corrections were relevant at macroscopic times, they could have 
an impact on the evolution of the Universe as a whole as well as on the structure formation because 
they imply a quantum correction to statistical fluctuations of energy density.

The existence of corrections to the classical expressions from relativistic 
kinetic theory in the cosmological metric was pointed out in refs.~\cite{fonarev1,fonarev2};
full expressions of energy density and pressure at global thermodynamic equilibrium in special 
curved space-times have been obtained, e.g. for the anti-De Sitter space-time \cite{allen,ambrus}. 
Herein, we derive a general expression of energy density and pressure for the free scalar field
in the spatially flat FLRW universe, where only a local thermodynamic equilibrium can be defined, 
showing that there are non-trivial corrections to the \eqref{cfs}. The obtained
expressions depend on the solutions of the Klein-Gordon equation in the FLRW metric, whose analytic
form essentially depends on the function $a(t)$. We will not work out the solutions for specific scale 
factor functions, which is a considerable problem on its own, deferring it to a paper dedicated
to this topic \cite{becarose2}.

\subsection*{Notation}

In this paper we use the natural units, with $\hbar=c=K=1$. The Minkowskian metric tensor $\eta$ is 
${\rm diag}(1,-1,-1,-1)$; for the Levi-Civita symbol we use the convention $\epsilon^{0123}=1$.\\ 
We will use the relativistic notation with repeated indices assumed to be saturated. Operators in Hilbert 
space will be denoted by a wide upper hat, e.g. $\widehat H$. Sometimes, scalar products of four-vectors
such as $V_\mu U^\mu$ will be denoted with a dot, i.e. $V \cdot U$. The Riemann tensor is defined as
$R^\mu_{\;\; \rho\lambda\nu} = \partial_\lambda \Gamma^\mu_{\rho\nu}- \partial_\nu \Gamma^\mu_{\rho\lambda}
+ \Gamma^\mu_{\lambda\sigma} \Gamma^\sigma_{\rho\nu}- \Gamma^\mu_{\nu\sigma} \Gamma^\sigma_{\rho\lambda}$

\begin{figure}
	\includegraphics[scale=0.5]{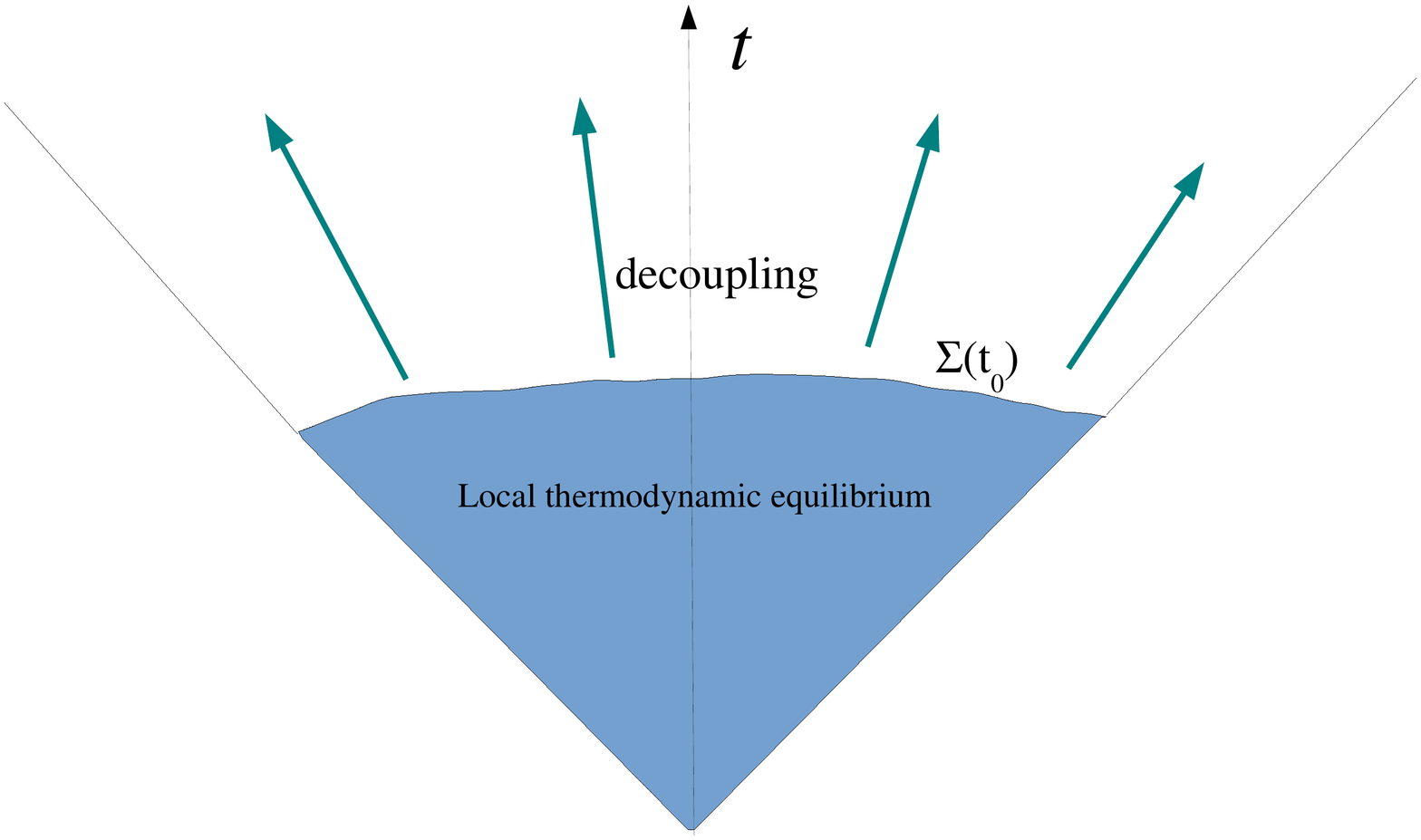}
	\caption{Schematic illustration of the evolution of a matter field as a function of cosmological
	time in the Friedman-Lema\^itre-Robertson-Walker universe. Until some time $t_0$ and the corresponding space-like
	hypersurface $\Sigma(t_0)$, the field is coupled to the	cosmological plasma, thereafter it decouples 
	and evolves freely. We take the approximation of an instantaneous cease of interactions, implying
	that decoupling - defined as the time where the expansion rate overcomes the interaction rate - 
	coincides with the freeze-out, defined as the time when interactions cease and the field becomes 
	effectively free.}
	\label{decoupling}
\end{figure}

\section{Local equilibrium and the expectation value of the stress-energy tensor}
\label{sect2}

The expectation value of the stress-energy tensor in curved space-time is a key ingredient for 
the solution of the semi-classical Einstein equation:
\be\label{einst}
 R_{\mu\nu} - \frac{1}{2} R g_{\mu\nu} = 8\pi G \langle \wT_{\mu\nu} \rangle_{\rm ren}
\ee
This depends, in the first place, on the quantum state of the Universe, which is in principle
unknown. The calculation of a renormalized value of $\wT^{\mu\nu}$ in an arbitrary pure quantum 
state is a difficult task, to which many theoretical efforts have been devoted since the '70s 
\cite{birrell}. Instead, we aim at calculating the stress-energy tensor originated from matter 
at thermodynamic equilibrium which thereafter decouples at some time in the evolution of the Universe, 
and whose form reduces to the equations \eqref{cfs} in the classical limit. In fact, our goal 
is just to obtain a proper quantum statistical extension of the expression \eqref{cfs}, derived 
by solving the free-streaming relativistic Boltzmann equation. 
Since the eq.~\eqref{cfs} originate from a local thermodynamic equilibrium at the time $t_0$, 
we consider the quantum density operator describing such a situation. In the general covariant 
formulation of quantum statistical mechanics \cite{zubarev,weert} this can be written as:
\be\label{rholeq}
   \wrho_{\rm LE} = \frac{1}{Z} \exp\left[ -\int_{\Sigma} \di\Sigma_\mu \wT^{\mu\nu} \beta_\nu \right]
\ee   
where $\di \Sigma_\mu = \di^3 x \; \sqrt{\gamma} n_\mu$, $\gamma$ being the determinant of the 
$3 \times 3$ metric tensor induced by $g$ onto the hypersurface $\Sigma$ and $n_\mu$ the unit vector 
perpendicular to $\Sigma$; $\beta_\nu = (1/T) u_\nu$ is the four-temperature vector, being 
$T$ the proper temperature and $u$ the four-velocity. The operator \eqref{rholeq} is obtained
by maximizing the entropy 
$$
  S = -\Tr (\wrho \log \wrho) 
$$
with the constraints of given energy and momentum densities \cite{betaframe}.

This form has been used to derive the so-called Kubo formulae of transport coefficients of 
relativistic hydrodynamics \cite{hosoya} and it is used for the modelling of the QCD plasma 
produced in relativistic nuclear collisions \cite{becazuba}.

The hypersurface $\Sigma$ is one where the condition of local equilibrium applies and is, to 
some extent, arbitrary; its choice effectively defines the initial state. For a system which 
is believed to maintain local equilibrium for some time, $\Sigma$ can be chosen to 
be the hypersurface of the latest time when local equilibrium applies, that is the decoupling time. 
In our approximation of a sudden interaction cease, as has been discussed in Section~\ref{sect1},
this coincides with the freeze-out hypersurface (see figure~\ref{decoupling}), which is the 
hypersurface $t=t_0$ in the FLRW coordinates. Therefore, throughout this paper, decoupling and 
freeze-out shall be considered synonymous expressions.

Henceforth, we will confine ourselves to the spacially flat FLRW metric, with $K=0$ in the eq.~\eqref{frw}. 
In this case, the operator $\wrho_{\rm LE}$ \eqref{rholeq}, on a constant time hypersurface, in the 
coordinates $t,{\bf x}$, takes the form:
\be\label{densop}
  \wrho_{\rm LE}(t) = \frac{1}{Z} \exp\left[ - a^3(t) \int \di^3 \x \; \wT^{00}(t,{\bf x})/T(t)\right]
\ee
Therefore, the operator:
\be\label{hami}
  \widehat H(t) = a^3(t) \int \di^3 \x \; \wT^{00}(t,{\bf x})
\ee
plays the role of an effective, time-dependent Hamiltonian, and the density operator \eqref{densop} 
can be written in the familiar form:
$$
 \wrho_{\rm LE}(t) = \frac{1}{Z} \e^{- \widehat H(t)/T(t)}
$$ 

It is important to stress that the thus-defined Hamiltonian operator is not conserved, i.e. 
depends on time. However, this operator is supposedly bounded from below, meaning 
that it has a Lowest Lying State (LLS) denoted as $\ket{0_{t}}$. Yet, since $\widehat{H}$ is not 
conserved, the LLS at the time $t$ is not the LLS of the Hamiltonian at some other time.

We now come to the problem of defining a finite value of the stress-energy tensor. 
The renormalization of the stress-energy tensor in a particular pure state, and 
especially the vacuum state, in a curved space-time is a well-known subject \cite{birrell}. 
In fact, we are interested here in the contribution of the excited (with respect 
to the vacuum) states to stress-energy tensor rather than the vacuum contribution. 
Therefore, we take as definition of mean stress-energy tensor the following:
\be\label{renset} 
  \langle T_{\mu\nu} \rangle = \Tr \left( \wrho_{\rm LE}(t_0) \wT^{\mu\nu} \right) 
   - \bra{0_{t_0}} \wT^{\mu\nu}(x) \ket{0_{t_0}}
\ee
where $t_0$ is the time of instantaneous decoupling/freeze-out. This formula implies that we 
take the density operator defining the state of the Universe as the fixed (as it should be in 
the Heisenberg representation) local thermodynamic equilibrium state at the decoupling/freeze-out 
time $t_0$. The eq.~\eqref{renset} is, therefore, the implementation, in the quantum statistical
framework, of the assumption that the mean value of the stress-energy tensor originates from a 
state of local thermodynamic equilibrium. It can be readily realized that in the limit $T(t_0) \to 0$, 
the right hand side of the \eqref{renset} vanishes and, consequently, any possible 
vacuum contribution to the stress-energy tensor \cite{glavan,glavan2} at zero temperature is 
simply subtracted away, alongside with all problems connected to the renormalization of 
the stress-energy tensor in vacuum including conformal and gravitational anomalies \cite{birrell}. 
Also, since both $\wrho_{\rm LE}(t_0)$ and $\ket{0_{t_0}}$ are not time-dependent quantum states, 
the stress-energy tensor defined in the equation \eqref{renset} fulfills the continuity equation 
$\nabla_\mu T^{\mu\nu}=0$ provided that the corresponding quantum operator does. 
Even though it may not be the complete right hand side of the Einstein equation \eqref{einst},
for some renormalized terms may be missing, the definition \eqref{renset} provides, 
as we shall see, a finite value of the stress-energy tensor. 

The above definition corresponds to the well-known one in flat space-time \cite{degroot}:
$$
 \langle T_{\mu\nu} \rangle = \Tr (\wrho :\wT^{\mu\nu}(x):)
$$
where the colon stands for normal ordering of creation and annihilation operators; the normal
ordering just implies the subtraction of the divergent vacuum term, which is unique in flat 
space-time. Indeed, normal ordering amounts to the subtraction of the vacuum expectation value 
of any operator which is quadratic in the creation and annihilation operators.
As it will be clear later, the definition \eqref{renset} gives rise to a finite value of 
the stress-energy tensor.

\section{The free scalar field in FLRW space-time}
\label{sect3}

The quantum stress-energy tensor operator appearing in \eqref{renset} depends on the considered quantum 
field. After freeze-out, interactions can be neglected and it is a very good approximation to 
regard the field as free. The simplest instance of a free quantum field is the real scalar field, 
whose equations of motion can be derived from the action:
\be\label{action}
 S = \frac{1}{2} \int \di^4 x \; \sqrt{-g} \, \left( \partial_\mu \wpsi \, \partial^\mu \wpsi - m^2 \wpsi^2
  + \xi R \wpsi^2 \right)
\ee
where $\xi$ is a real scalar coefficient proportional to the curvature-field coupling; noteworthy
cases are $\xi=0$, the so-called minimal coupling and $\xi=1/6$, the so-called conformal coupling
which makes the action conformally invariant in the massless case. The equation of motion of the
field are the Klein-Gordon equations in curved spacetime:
\be\label{kg}
 ( \Box + m^2 - \xi R ) \wpsi = 0
\ee
with $\Box = \nabla_\mu \nabla^\mu$. The stress-energy tensor can be obtained from the action by
taking the functional derivative with respect to the metric $g_{\mu\nu}$ and reads:
\be\label{setkg}
 \wT_{\mu\nu} = \nabla_\mu \wpsi \nabla_\nu \wpsi - \frac{1}{2} g_{\mu\nu} ( \nabla \wpsi \cdot \nabla \wpsi
  - m^2 \wpsi^2 ) + \xi ( G_{\mu\nu}  + g_{\mu\nu} \Box - \nabla_\mu \nabla_\nu ) \wpsi^2
\ee
where $G_{\mu\nu} = R_{\mu\nu} - (1/2) R g_{\mu\nu}$ is the Einstein tensor.
The solution of the Klein-Gordon equation in the FLRW metric has been extensively studied in literature 
\cite{mukhanov,birrell}. This is easier to obtain in conformal coordinates, where the $t$ coordinate is
replaced by:
\be\label{conftime}
  \eta = \int_{t_0}^t \frac{\di t}{a(t)}
\ee
so that $\eta=0$ at the freeze-out and, for the spacially flat metric, with spacial Cartesian coordinates:
$$
 \di s^2 = a(\eta)^2 (\di \eta^2 - \di x^2 - \di y^2 -\di z^2 )
$$
Like for the Minkowski space-time, the field can be expanded in plane waves. By using the notations 
${\bf k} = (k_x,k_y,k_z)$, by ${\bf k} \cdot{\bf x} = k_x x + k_y y + k_z z$ and 
$\kk^2 = k_x^2 + k_y^2 +k_z^2$, one has: 
\be\label{field}
 \wchi \equiv a \wpsi  = \frac{1}{(2\pi)^\frac{3}{2}} \int \di^3 \kk \;( v_{k} \e^{\ii {\bf k}\cdot{\bf x}}
 \wa{({\bf k})} + v^*_{k} \e^{-\ii {\bf k}\cdot{\bf x}} \wad{({\bf k})})
\ee
where the $\wa{({\bf k})}$ are the annihilation operators \footnote{The annihilation operators
$\wa{({\bf k})}$ should not be confused with the scale factor $a$; they appear with an upper hat
throughout the paper}. It will become clear later on, through the comparison with the classical 
expressions \eqref{cfs}, that ${\bf k} = (k_x,k_y,k_z)$ are the covariant components of the 
momentum of a particle in the comoving cosmological coordinates. 

The eigenfunctions $v_k$ fulfill the equation, ensuing from the Klein-Gordon equation \eqref{kg}:
\be\label{kg2}
   v''_k + \Omega^2_k v_k = 0  \qquad\qquad \Omega^2_k = \kk^2 + m^2 a^2 - (1-6\xi) \frac{a^{''}}{a}
\ee
where the prime denotes a derivative with respect to $\eta$. As it is implied by their subscript, 
they only depend on the magnitude of the vector ${\bf k}$, which is a consequence of the dependence 
of $\Omega$ on it. For the creation and annihilation operators to fulfill the commutation relation:
\be\label{commrel}
  [\wa{({\bf k})},\wad{({\bf k}^\prime)}] = \delta^3({\bf k}-{\bf k}^\prime)
\ee
the eigenfunctions $v_k$ must be normalized so that their Wronskian coincides with the imaginary 
unit, that is:
$$
 W [v_k] = v_k v^{*'}_k - v'_k v^{*}_k = \ii
$$

A very well known issue in quantum field theory in curved space-time is the ambiguity of the 
vacuum state, that is the state which is annihilated by all the operators $\wa{({\bf k})}$. The
ambiguity is related to the choice of the eigenfunctions $v_k$, which is arbitrary to a large 
extent. If $v_k$ is a set of eigenfunctions, another set can be generated by taking a linear
superposition of $v_k$ and its complex conjugate:
$$
 w_k = A_k v_k + B_k v^*_k
$$
with $|A_k|^2 - |B_k|^2 = 1$ to maintain the same value of the Wronskian. This new set of 
eigenfunctions corresponds to another expansion of the field \eqref{field} with a new set of creation
and annihilation operators which are related to the original ones through a Bogoliubov transformation:
\be\label{bogo}
  \wc{({\bf k})} = A^*_k \wa{({\bf k})} - B^*_k \wad{(-{\bf k})}
\ee
It appears as a straightforward consequence of the \eqref{bogo} that the state which is annihilated
by all the $\wa{}$'s is not annihilated by the $\wc{}$'s and vice-versa, hence the vacua of $\wa{}$'s
and $\wc{}$'s differ. 

\subsection{The Hamiltonian and the choice of the vacuum}

Following the arguments of Section \ref{sect2}, our goal is to determine the
set of eigenfunctions and their associated annihilation operators having as vacuum state the 
instantaneous LLS \cite{mukhanov} of the Hamiltonian \eqref{hami} at the time $t_0$ or $\eta = 0$.
The calculation of the Hamiltonian requires the integration of $\wT_{00}$ (defined as the time-time
component in Cartesian coordinates). By using the eq. \eqref{setkg} with the field $\wchi = a \wpsi$,
at general time $t$ or its corresponding $\eta$, this turns out to be (see Appendix A):
\be\label{tzero}
 \wT_{00} = \frac{1}{2a^4} \left[ \wchi^{\prime 2} +\vec{\nabla}\widehat{\chi}\cdot\vec{\nabla}
 \widehat{\chi} + m^2 a^2 \wchi^2 + (1-6\xi) \left( \frac{a^{\prime 2}}{a^2} \wchi^2 - 
 2 \frac{a'}{a} \wchi^\prime \wchi \right) -4\xi(\widehat{\chi}\vec{\nabla}^2\widehat{\chi}
 +\vec{\nabla}\widehat{\chi}\cdot\vec{\nabla}\widehat{\chi}) \right]
\ee 
where $\vec{\nabla}$ stands for the array $(\partial_x,\partial_y,\partial_z)$ and $\vec{\nabla}^2
= \partial_x^2 + \partial_y^2 + \partial_z^2$.
By plugging the field expansion \eqref{field} into the \eqref{tzero}, and after the integration 
in $\di^3 \x$, the equation \eqref{hami} yields (see Appendix B):
\be\label{hami3}
 \widehat H (\eta) = \frac{1}{2a(\eta)} \int \di^3 \kk \; \omega_k(\eta) \left[ K_k(\eta) 
 \left( \wad{({\bf k})} \wa{({\bf k})} + \wa{({\bf k})} \wad{({\bf k})} \right) + 
 \Lambda_k(\eta) \wa{({\bf k})} \wa{(-{\bf k})} +
 \Lambda^*_k(\eta) \wad{(-{\bf k})} \wad{({\bf k})} \right]
\ee
with
\be\label{funz}
\begin{split}
\omega^2_k(\eta) &= \kk^2 + m^2 a^2 + (1-6\xi) \frac{a^{\prime 2}}{a^2} \\
K_k(\eta) &= \frac{1}{\omega_k} \left[ |v'_k|^2 + \omega^2_k |v_k|^2 -2 (1-6\xi) \frac{a'}{a} 
{\rm Re} (v'_k v^*_k)\right] \\
\Lambda_k(\eta) &= \frac{1}{\omega_k} \left[ v^{\prime 2}_k + \omega^2_k v_k^2 -2 (1-6\xi) \frac{a'}{a} 
v'_k v_k \right]
\end{split}
\ee
Thus, in order for $\ket{0}$, which is the the vacuum state of the operators $\wa{({\bf k})}$,
to be at the same time the LLS of the Hamiltonian $\widehat{H}(t_0)$, i.e. $\ket{0_0}$,
keeping in mind that $t_0$ corresponds to $\eta=0$ according to the eq.~\eqref{conftime}, one needs
to have, from \eqref{hami3}:
$$
 \widehat H (t_0) \ket{0} = \frac{1}{2 a} \int \di^3 \kk \; \omega_k(0) \left[ K^*_k(0) \ket{0}
 + \Lambda^*_k(0) \wad{(-{\bf k})} \wad{({\bf k})} \ket{0}\right] = E_0 \ket{0} = E_0 \ket{0_0}
$$
where we have used the commutation relations \eqref{commrel} and $E_0$ is the lowest lying eigenvalue. 
This is achieved if $\Lambda_k(0) = 0$ and $K_k(0)$, which is real, is minimal. It can be shown that 
both these conditions occur if:
\be\label{incond}
 v_k(0) = \frac{1}{\sqrt{2 \omega_\xi(0,k)}} \qquad \qquad v'_k(0) 
 = -\frac{\ii}{2 v_k(0)} + (1-6\xi) \frac{a'(0)}{a(0)} v_k(0)
\ee
with
\be\label{omegaxi}
 \omega_\xi(\eta,k) = \sqrt{\omega^2_k(\eta) - (1-6\xi)^2 \frac{a^{\prime 2}}{a^2}} = 
 \sqrt{\kk^2 + m^2 a^2 + 6\xi (1-6\xi) \frac{a^{\prime 2}}{a^2}}
\ee
with an overall phase factor of $v$ which remains undetermined; this is not an issue
because a phase factor can always be absorbed in a redefinition of the creation and annihilation
operators. In fact, the very existence of the solution \eqref{incond} requires:
$$
 \omega_k^2(0) > (1 - 6\xi)^2 \frac{a^{\prime 2}(0)}{a^2(0)}  \implies 
 \kk^2 + m^2 a^2(0) + 6\xi(1-6\xi)  \frac{a^{\prime 2}(0)}{a^2(0)} > 0
$$
which is certainly true if $0 \le \xi \le 1/6$; otherwise, this depends on the relative 
magnitude of the classical term $\kk^2+m^2a^2(0)$ and the other term which is a quantum correction 
proportional to $a^{\prime 2}(0)$. By plugging the eq.~\eqref{incond} into the eq.~\eqref{funz} 
one obtains:
\be\label{kzero}
  K_k(0) = \frac{\omega_\xi(0,k)}{\omega_k(0)}
\ee
This is indeed the minimal value of $K_k(0)$ for it can be readily checked that, with the definitions
\eqref{funz}, the following constraint is fulfilled:
\be\label{klcons}
    K_k^2(\eta)-\left|\Lambda_k(\eta)\right|^2=
    \frac{\omega^2_\xi(\eta,k)}{\omega_k^2(\eta)}
\ee
so that the minimum of $K_k(0)$ is achieved when $\Lambda_k(0)=0$.

With the previous results, the lowest lying eigenvalue can be obtained:
$$
 E_0 = \frac{1}{2 a} \int \di^3 \kk \; \omega_\xi(0,k)
$$
which is divergent. This is an expected result in full agreement with the flat space-time limit, 
$a=1 \;\; \omega_\xi = \sqrt{\kk^2+m^2}$, where the constant $E_0$ corresponds to the zero-point 
energy of the free field modes. Moreover, as $\Lambda_k(0)=0$, the Hamiltonian in eq.~\eqref{hami3} 
at $\eta=0$ becomes diagonal in the quadratic form $\wad{({\bf k})}\wa{({\bf k})}$:
\be\label{hami4}
\widehat H (0) = \int \di^3 \kk \; \omega_\xi(0,k) \left( \wad{({\bf k})} 
\wa{({\bf k})} + \frac{1}{2} \right)
\ee
keeping in mind that $a(0) = 1$. 

The divergences will be removed by subtracting the vacuum expectation values, according to the 
prescription \eqref{renset}, which amounts to take normal ordering in the flat space-time case as 
discussed in Section~\ref{sect2}.

\section{Energy density and pressure}
\label{sect4}

We are now in a position to evaluate the vacuum-subtracted expectation value of the 
stress-energy tensor by using the formula \eqref{renset} at $\eta=0$: 
\begin{equation}\label{renset2}
T_{\mu\nu}(x)= \Tr \left( \widehat{\rho}_{\rm LE}(0) \widehat{T}_{\mu\nu}(x)\right) - \bra{0_0} 
\wT_{\mu\nu}(x) \ket{0_0} = \frac{1}{Z}\Tr \left( \exp\left[-\frac{\widehat{H}(0)}{T(0)}\right]
\widehat{T}_{\mu\nu}(x)\right) - \bra{0_0} \wT_{\mu\nu}(x) \ket{0_0}
\end{equation}
For this purpose we shall need the expectation values of quadratic combinations of creation
and annihilation operators such as:
$$
 \langle \wad{({\bf k})} \wa{({\bf k}')} \rangle = \Tr \left( \exp\left[-\frac{\widehat{H}(0)}{T(0)}\right]
 \wad{({\bf k})} \wa{({\bf k}')} \right)
$$
Such forms are straightforward to work out, as the Hamiltonian is diagonal in the creation 
and annihilation operators, see eq.~\eqref{hami4}. By using the traditional methods of thermal
field theory, one readily finds:
\begin{equation}\label{valasp}
\begin{split}
\left<\widehat{a}(\mathbf{k})\widehat{a}(\mathbf{k}')\right>&=
\left<\widehat{a}^\dagger(\mathbf{k})\widehat{a}^\dagger(\mathbf{k}')\right>=0,\\
\left<\widehat{a}(\mathbf{k})\widehat{a}^\dagger(\mathbf{k}')\right>&=
n_B\left( \frac{\omega_\xi(0,k)}{T(0)}\right) \delta^3(\mathbf{k}-\mathbf{k}'),\\
\left<\widehat{a}^\dagger(\mathbf{k})\widehat{a}(\mathbf{k}')\right>&=
\left[ n_B\left(\frac{\omega_\xi(0,k)}{T(0)}\right)+1 \right] \delta^3(\mathbf{k}-\mathbf{k}').
\end{split}
\end{equation}
where $n_B$ is the Bose-Einstein distribution:
\be\label{bose}
 n_B\left( \frac{\omega_\xi(0,k)}{T(0)}\right) \equiv 
 \frac{1}{\exp\left[\omega_\xi(0,k)/T(0) \right]-1},
\ee
and $\omega_\xi$ is given by \eqref{omegaxi}.

We begin with the time-time component of the stress-energy tensor in the Cartesian coordinates. 
Plugging the field expansion \eqref{field} into the \eqref{tzero} we have
\begin{equation*}
    \begin{split}
    \Tr\left(\widehat{\rho}_{\rm LE}(0)\widehat{T}_{00}(x)\right)&
    =\frac{1}{2a^4(2\pi)^3}\int \di^3 \kk \int \di^3 \kk' \; \\
&\times\left\{\e^{-\ii\mathbf{x}\cdot(\mathbf{k}+\mathbf{k}')}\left[v^{*\prime}_k v^{*\prime}_{k'}+
\left( \mathbf{k}'\cdot\mathbf{k}+m^2a^2+(1-6\xi)\frac{a'^2}{a^2} \right) v^*_k v^*_{k'}-
\left(1-6\xi\right)\frac{a'}{a} v^{*'}_kv^*_{k'} \right]
\left<\widehat{a}^\dagger(\mathbf{k})\widehat{a}^\dagger(\mathbf{k}')\right>\right.\\
&  +\e^{\ii\mathbf{x}(\mathbf{k}+\mathbf{k}')}\left[v'_k v'_{k'}+\left( \mathbf{k}'\cdot
    \mathbf{k}+m^2a^2+(1-6\xi)\frac{a'^2}{a^2} \right) v_kv_{k'}-\left(1-6\xi\right)\frac{a'}{a}v'_kv_{k'}\right]
    \left<\widehat{a}(\mathbf{k} )\widehat{a}(\mathbf{k}')\right>\\
    &+\e^{\ii\mathbf{x}\cdot(-\mathbf{k}+\mathbf{k}')}\left[v'_k v^{*\prime}_{k'}+
    \left( \mathbf{k}'\cdot\mathbf{k}+m^2a^2+(1-6\xi)\frac{a'^2}{a^2} \right)
    v_k v^*_{k'}-\left(1-6\xi\right)\frac{a'}{a}v'_kv^*_{k'} \right]
    \left<\widehat{a}^\dagger(\mathbf{k})\widehat{a}(\mathbf{k}')\right>\\
    &+\left.\e^{\ii\mathbf{x}\cdot(\mathbf{k}-\mathbf{k}')}\left[v^{*\prime}_kv'_{k'}+
   \left( \mathbf{k}'\cdot\mathbf{k}+m^2a^2+(1-6\xi)\frac{a'^2}{a^2} \right) v^*_kv_{k'}-
   \left(1-6\xi\right)\frac{a'}{a}v^{*\prime}_kv_{k'}\right]
    \left<\widehat{a}(\mathbf{k})\widehat{a}^\dagger(\mathbf{k}')\right>\right\},
    \end{split}
\end{equation*}
which reduces to, by using the equations \eqref{valasp} and \eqref{funz}:
\be\label{unrent00}
\begin{split}
\Tr\left(\widehat{\rho}_{\rm LE}(0)\widehat{T}_{00}(x)\right) &=\frac{1}{a^4(\eta)(2\pi)^3}
 \int \di^3 \kk \;\left[\left|v'_k(\eta)\right|^2+\omega^2_k(\eta)
\left|v_k(\eta)\right|^2-2(1-6\xi)\frac{a'(\eta)}{a(\eta)}\mbox{Re}(v'_k(\eta)v^*_k(\eta))\right] \\
& \times \left[n_B\left( \frac{\omega_\xi(0,k)}{T(0)} \right)+\frac{1}{2}\right] \\ 
& =\frac{1}{a^4(\eta)(2\pi)^3}\int \di^3\kk\; \omega_k(\eta) K_k(\eta)
\left[ n_B \left( \frac{\omega_\xi(0,k)}{T(0)}\right) +\frac{1}{2}\right],
\end{split}
\ee
where time dependencies have been spelled out for the sake of clarity.

The vacuum expectation value needs now to be evaluated and subtracted, according to the
definition \eqref{renset2}. As the LLS of the Hamiltonian \eqref{hami4} is the vacuum $\ket{0_0}$, 
this is simply obtained by taking the limit $T(0) \to 0$ in the equation \eqref{unrent00}:
\begin{equation}
   \bra{0_0} \wT_{00}(x) \ket{0_0} = \lim_{T(0)\to 0} 
   \Tr \left(\wrho_{\rm LE}(0) \widehat{T}_{00}(x) \right) = 
    \frac{1}{2 a^4(\eta)(2\pi)^3} \int \di^3 \kk\; \omega_k(\eta)K_k(\eta)
\end{equation}
Therefore, altogether, the vacuum-subtracted value of $T_{00}(x)$ is obtained by removing the term 
proportional to $1/2$ in the equation \eqref{unrent00}:
\be\label{t00f}
T_{00}(x) = \frac{1}{a^4(\eta)(2\pi)^3}\int \di^3\kk\; \omega_k(\eta) K_k(\eta)
n_B \left( \frac{\omega_\xi(0,k)}{T(0)}\right)
\ee

The finite, vacuum-subtracted, expectation values of the space-space components of the stress-energy 
tensor can be obtained likewise. It is convenient to take advantage of the isotropy of the metric 
and calculate the expectation value of each diagonal component as 1/3 of that of the spacial trace 
(in Cartesian coordinates) namely:
$$
\Tr (\wrho_{\rm LE}(0) \, \wT_{jj} ) = 
\frac{1}{3} \Tr \left( \wrho_{\rm LE}(0) \sum_{i=1}^3 \wT_{ii} \right) \qquad \qquad j=1,2,3
$$
It can be shown, by using the definition \eqref{setkg} (see Appendix A):
\begin{equation}\label{tii}
\begin{split}
\sum_{i=1}^3 \widehat{T}_{ii}=\frac{1}{2a^2}&\left[\widehat{\chi}'^2(3-12\xi)+
\vec{\nabla}\widehat{\chi}\cdot\vec{\nabla}\widehat{\chi}(12\xi-1)- 6\widehat{\chi}'\widehat{\chi}
\frac{a'}{a}(1-6\xi)-4\xi(\widehat{\chi}\vec{\nabla}^2\widehat{\chi}+\vec{\nabla}\widehat{\chi}
\cdot\vec{\nabla}\widehat{\chi})+\right.\\
&\left.+\widehat{\chi}^2 \left( 3 m^2a^2(4\xi-1) + 3\frac{a'^2}{a^2}(1-6\xi)+\frac{a''}{a} 12 \xi(6\xi-1)
\right)\right].
\end{split}
\end{equation}
By plugging the field expansion \eqref{field} in the \eqref{tii} one obtains:
\begin{equation*}
    \begin{split}
     \mbox{Tr}\left(\widehat{\rho}(0)\widehat{T}_{jj}(x)\right)&=\frac{1}{3}
     \mbox{Tr}\left(\widehat{\rho}(0)\sum^3_{i=1}\widehat{T}_{ii}\right)=\frac{1}{a^2(\eta)\left(2\pi\right)^3}
     \int \di^3 \kk \; \left\{\left(1-4\xi\right)\left|v'_k\right|^2-2\left(1-6\xi\right)
     \left(v'_kv^*_k+v^{*'}_kv_k\right)\right.\\
     &+\left.\left[\left(4\xi-\frac{1}{3}\right)\kk^2+3m^2a^2\left(4\xi-1\right)+
     \left(1-6\xi\right)\frac{a'^2}{a^2}-4\xi\left(1-6\xi\right)\frac{a''}{a}\right]\left|v_k\right|^2\right\}\\
     &\times\left(n_B\left(\frac{\omega_\xi\left(0,k\right)}{T(0)}\right)+\frac{1}{2}\right),
    \end{split}
\end{equation*}
where the quantities inside the braces are calculated at a general time $\eta$.
The above expression above can be written in a compact form:
\be\label{tii2}
 \Tr \left( \wrho_{\rm LE}(0) \widehat{T}_{jj}(x)\right) = \frac{1}{a^2(2\pi)^3} \int \di^3 
 \kk \; \omega_k(\eta) \, \Gamma_k(\eta) \left[ n_B\left( \frac{\omega_\xi(0,k)}{T(0)}\right) 
  +\frac{1}{2}\right]
\ee
where:
\be\label{funzgamma}
 \Gamma_k(\eta)=\frac{1}{\omega_k(\eta)}\left[(1-4\xi)\left|v'_k(\eta)\right|^2+\frac{1}{3} \gamma_k(\eta)
 \left|v_k(\eta)\right|^2-2(1-6\xi)\frac{a'(\eta)}{a(\eta)}\mbox{Re}(v'_k(\eta)v^*_k(\eta))\right]
\end{equation}
with:
\be\label{gamma}
   \gamma_k(\eta)=\left( 12\xi-1 \right) \kk^2 + 3 (4\xi-1) m^2 a^2(\eta)+
   3 (1-6\xi)\frac{a'^2(\eta)}{a^2(\eta)} - 12 \xi(1-6\xi)\frac{a''(\eta)}{a(\eta)}
\ee
By using the initial conditions \eqref{incond}, the \eqref{funzgamma} at the decoupling time
$\eta=0$ becomes:
\begin{equation}\label{gamma0}
\begin{split}
\Gamma_k(0)=\frac{1}{\omega_k(0)\omega_\xi(0,k)}
\left[\frac{\kk^2}{3}-2\xi\left(1-6\xi\right)\left(a''(0)+a'^2(0)\right)\right].
\end{split}
\end{equation}
with $\omega_{\xi}(\eta,k)$ given by \eqref{omegaxi} and $\omega_k$ by \eqref{funz}.
Like for the $T_{00}$ component, the vacuum-subtracted value can be obtained by 
subtracting, from the equation \eqref{tii2} its limit for $T(0) \to 0$, yielding:
\be\label{tiif}
 T_{jj}(x) = \frac{1}{a^2(2\pi)^3} \int \di^3 
 \kk \; \omega_k(\eta) \, \Gamma_k(\eta) n_B\left( \frac{\omega_\xi(0,k)}{T(0)}\right)\qquad\qquad j=1,2,3
\ee

It is now straightforward to extract, from the equations \eqref{t00f} and \eqref{tiif}, the energy
density and pressure in the equation \eqref{set}:
\begin{equation}\label{EeP}
\begin{split}
 \varepsilon(\eta)&=\frac{1}{a^4(\eta)(2\pi)^3}\int \di^3 \kk \;  
  \omega_k(\eta) K_k(\eta) n_B \left(\frac{\omega_\xi(0,k)}{T(0)} \right),\\
  p(\eta)&=\frac{1}{a^4(\eta)(2\pi)^3} \int \di^3 \kk \; \omega_k(\eta)\Gamma_k(\eta)
   n_B \left( \frac{\omega_\xi(0,k)}{T(0)} \right).
\end{split}
\end{equation}
The equations \eqref{EeP} are the main result of this work. It can be shown that the functions 
in equation \eqref{EeP} fulfill the covariant conservation of the stress-energy tensor 
(see appendix \ref{a4}).

It can be seen that if $\xi=0$ or $\xi=1/6$, one has:
$$
  \omega_\xi(\eta,k) = \sqrt{\kk^2 + m^2 a^2(\eta)}
$$
Furthermore, if the coefficients $K_k(\eta)$ and $\Gamma_k(\eta)$, defined in the equation 
\eqref{funz}, are such that:
\be\label{classical}
\begin{split}
 & \omega_k(\eta) K_k(\eta) = \sqrt{\kk^2 + m^2 a^2(\eta)} \\ \nonumber
 & \omega_k(\eta) \Gamma_k(\eta) = \frac{\kk^2}{3\sqrt{\kk^2+m^2a^2(\eta)}}
\end{split}
\ee
then the expressions \eqref{EeP}, taking the \eqref{bose} into account, precisely reproduce the classical 
equations \eqref{cfs}, which certifies the physical interpretation of $\bf k$ in the field 
expansion \eqref{field} as the covariant components of the particle momentum. 

It can be readily checked that if $a(\eta)={\rm constant}$, by using the definitions of $K_k(\eta)$, 
$\Gamma_k(\eta)$ and taking into account that the solution of the Klein-Gordon equation \eqref{kg2} 
with the initial conditions \eqref{incond} is the familiar
$$
v=\frac{\e^{-\ii \eta \omega_k}}{\sqrt{2 \omega_k}}
$$
with $\omega_k=\sqrt{\kk^2+m^2 a^2}$, the equations \eqref{classical} are fulfilled.
However, for a general function $a(\eta)$ the above equalities \eqref{classical} are not fulfilled, 
so that the equations \eqref{EeP} involve non-trivial corrections to the classical free-streaming 
expressions that we will examine in the next Section.
Those corrections make the equation of state, meant as the relation between energy density and 
pressure, more complicated than that predicted by the relativistic Boltzmann equation, which is 
encoded in the eq.~\eqref{cfs}. Particularly, the equation of state will depend on the solutions of the
Klein-Gordon equation in FLRW space-time for some given scale factor evolution function $a(\eta)$
because so do $K_k(\eta)$ and $\Gamma_k(\eta)$ in eq.~\eqref{funz}.
Thus, in principle, the equation of state at some given time $\eta$ depends functionally on the scale 
factor, that is on the entire evolution of the metric from the decoupling onwards.

\section{Corrections to the energy density and pressure}
\label{sect5}

The \eqref{EeP} depend on the parameter $\xi$ gauging the coupling between curvature and field
in the action \eqref{action}. Amongst the corrections to the classical free streaming solution
\eqref{cfs}, the replacement of $\sqrt{\kk^2+m^2 a^2}$ with $\omega_\xi$ in eq. \eqref{omegaxi}
is certainly an evident one. The energy $\omega_\xi(0,k)$ in the Bose-Einstein distribution includes an extra 
term proportional to $(a'/a)^2$ (see eq \eqref{omegaxi}) which is clearly of quantum origin as it requires
an $\hbar^2$ in standard units to be added to an energy squared. However, this term vanishes 
for the two special values $\xi=0$ and $\xi=1/6$, corresponding to the minimal and conformal coupling.
These two cases are indeed the best known and physically motivated, thus they deserve a special attention.

\subsection{Conformal coupling: $\xi = 1/6$}

The conformal coupling is the easiest to handle as the previous expressions simplify considerably. 
First, from eqs.~\eqref{funz} and \eqref{omegaxi}:
\be\label{omegas}
  \omega_k(\eta)=\omega_\xi(\eta,k)= \Omega_k(\eta) = \sqrt{\kk^2 + m^2 a^2}
\ee
and, from the equation \eqref{gamma}:
$$
\gamma_k(\eta)= \kk^2 - m^2 a^2(\eta)
$$
Furthermore, from the equations \eqref{funz} and \eqref{funzgamma}:
\be\label{funzconf}
    \begin{split}
K_k(\eta) &= \frac{1}{\omega_k} \left( |v'_k|^2 + \omega^2_k |v_k|^2 \right) \\ 
 \Gamma_k(\eta) &= \frac{1}{3\omega_k} \left( |v'_k|^2 + \gamma_k |v_k|^2 \right)        
    \end{split}
\ee
It is interesting to point out that those expressions, once used in the eqs. \eqref{EeP}, imply 
the remarkable inequalities:
$$
    \varepsilon \pm 3 p > 0 
$$
which are relevant for the cosmological Friedman equations. 

By comparing the eq.~\eqref{EeP} with the eq.~\eqref{cfs}, and using the \eqref{omegas} we can
identify the corrections to the classical free-streaming solution as:
\begin{equation}\label{quantumcor}
    \begin{split}
 \Delta\varepsilon(\eta)&=\frac{1}{a^4(\eta)(2\pi)^3}\int \di^3 \kk \; \omega_k(\eta)
 \left[ K_k(\eta) -1 \right] n_B \left( \frac{\omega_k(0)}{T(0)} 
 \right),\\
 \Delta p(\eta)&=\frac{1}{a^4(\eta)(2\pi)^3}\int \di^3 \kk \; \omega_k(\eta)
 \left[\Gamma_k(\eta)-\frac{\kk^2}{3 \omega^2_k(\eta)}\right]
n_B \left( \frac{\omega_k(0)}{T(0)} \right).
    \end{split}
\end{equation}

By using the initial conditions \eqref{incond}, with $\xi=1/6$, in the eq.~\eqref{funzconf} one 
has, being $a(0)=1$:
$$
 K_k(0)- 1 = 0 \qquad
 \Gamma_k(0) -\frac{\kk^2}{3(\kk^2+m^2)} = 
 \frac{\kk^2}{3\omega^2_k(0)} -\frac{\kk^2}{3\omega^2_k(0)} = 0
$$ 
Consequently, both $\Delta \varepsilon(0)$ and $\Delta p(0)$ vanish in the equation \eqref{quantumcor}.
The important question is whether the quantum corrections become comparable to the main classical 
free-streaming term \eqref{cfs} at some late time $\eta$. The answer to this question requires the 
solution of the Klein-Gordon equation \eqref{kg2}, which in turn requires the knowledge of the function 
$a(\eta)$, what goes beyond the scope of the present work and it will the main subject
of a forthcoming study \cite{becarose2} as it has been mentioned in Section \ref{sect1}.
Nonetheless, it is possible to gain some insight on the possible impact of $\Delta \varepsilon$
and $\Delta p$ by studying the behaviour of the square brackets in eq.~\eqref{quantumcor} near the 
freeze-out time $\eta=0$. Particularly, by using again the initial conditions \eqref{incond} and the 
Klein-Gordon equation \eqref{kg2}, it can be shown that:
\begin{equation*}
    \begin{split}
 & \frac{\di}{\di \eta}\left( K_k(\eta) - 1 \right) \Big|_{\eta=0} = 0 \qquad\qquad 
  \frac{\di^2}{\di \eta^2}\left( K_k(\eta) - 1 \right) \Big|_{\eta=0} = \frac{m^4}{(m^2+\kk^2)^2} a'^2(0) \\
 & \frac{\di}{\di \eta}\left( \Gamma_k(\eta) - \frac{\kk^2}{3\omega_k^2(\eta)} \right) \Big|_{\eta=0} = 
 - \frac{m^4}{3(\kk^2+m^2)^2} a'(0)
    \end{split}
\end{equation*}
Therefore, we conclude that, for a small conformal time $\Delta\eta$ after the freeze-out:
\be\label{smalltimes1}
  \left( K_k(\eta) - 1 \right) \simeq \frac{1}{2} \frac{m^4}{(m^2+\kk^2)^2} a'^2(0) (\Delta\eta)^2
  \qquad
  \left( \Gamma_k(\eta) - \frac{\kk^2}{3\omega_k^2(\eta)} \right) \simeq - \frac{m^4}{3(\kk^2+m^2)^2} a'(0) 
  \Delta\eta
\ee
hence the expansion of the FLRW space-time has an opposite effect on the quantum corrections of
energy density and pressure right after the freeze-out, that is:
$$
\Delta \varepsilon > 0  \qquad\qquad \Delta p < 0
$$
Whether these trends continue at long times after the freeze-out and make the corrections to
the classical term comparable or even predominant over it cannot be shown without solving the 
Klein-Gordon equations \eqref{kg2}, as has been mentioned. 

Finally, we note that in the $m=0$ case, with $\xi=1/6$, the Klein-Gordon equation \eqref{kg2} can
be explicitly solved and the solution fulfilling the initial conditions \eqref{incond} is the familiar 
exponential function:
$$
 v_k = \frac{1}{2\kk} \e^{-i \kk \eta}
$$
whence $K_k(\eta) = 1$ and $\Gamma_k(\eta) = 1/3$. Furthermore:
$$
    \Delta\varepsilon(\eta)=\Delta p(\eta)=0 \qquad\forall\eta
$$
This is a consequence of the known fact that, in the massless case with conformal coupling, the 
operator \eqref{rholeq} is a global thermodynamic equilibrium state because the trace of the stress-energy
tensor vanishes.

\subsection{Minimal coupling: $\xi = 0$}

In this case we have:
\be\label{omegas2}
    \omega_k(\eta)=\sqrt{\kk^2+m^2a^2+\frac{a'^2}{a^2}},\quad\Omega_k(\eta)=
    \sqrt{\kk^2+m^2a^2-\frac{a''}{a}},\quad\omega_\xi\left(\eta,k\right)=\sqrt{\kk^2+m^2a^2}
\ee
With these expressions, the corrections $\Delta \varepsilon$ and $\Delta p$ can be written 
as follows:
\begin{equation}\label{quantumcor2}
    \begin{split}
 \Delta\varepsilon(\eta)&=\frac{1}{a^4(\eta)(2\pi)^3}\int \di^3 \kk \; \left[ \omega_k(\eta)
  K_k(\eta) - \sqrt{\kk^2 + m^2 a^2(\eta)}\right] n_B \left( \frac{\omega_k(0)}{T(0)} 
 \right),\\
 \Delta p(\eta)&=\frac{1}{a^4(\eta)(2\pi)^3}\int \di^3 \kk \; 
 \left[\omega_k(\eta) \Gamma_k(\eta)-\frac{\kk^2}{3\sqrt{\kk^2 + m^2 a^2(\eta)} }\right]
n_B \left( \frac{\omega_k(0)}{T(0)} \right).
    \end{split}
\end{equation}
The function $\gamma_k$ in the \eqref{gamma} becomes:
\begin{equation*}
    \gamma_k(\eta)=-\kk^2- 3 m^2 a^2 + 3\frac{a'^2}{a^2},
\end{equation*}
and the functions \eqref{funz}, \eqref{funzgamma} reduce to
\begin{equation*}
    \begin{split}
        K_k(\eta)&=\frac{1}{\omega_k}\left[\left|v'_k\right|^2+\omega^2_k\left|v_k\right|^2-
        \frac{2a'}{a}\mbox{Re}\left(v'_kv^*_k\right)\right],\\
        \Gamma_k(\eta)&=\frac{1}{\omega_k}\left[\left|v'_k\right|^2+\frac{1}{3}
        \gamma_k\left|v_k\right|^2-\frac{2a'}{a}\mbox{Re}\left(v'_kv^*_k\right)\right].
    \end{split}
\end{equation*}
The only inequality which is implied by the above expressions is:
$$
  \varepsilon > p
$$
Nevertheless, like in the conformal case, the corrections at the freeze-out time $\Delta \varepsilon(0)$ 
and $\Delta p(0)$ vanish. This can be shown by using the equation \eqref{kzero} for the 
energy density, keeping in mind the \eqref{omegas2}, and the \eqref{gamma0} for the pressure.

As done in the conformal coupling case we can gain some insight on the behaviour of 
$\Delta\varepsilon$ and $\Delta p$ near the freeze-out studying the derivatives of the functions 
$K_k$ and $\Gamma_k$. Using the initial conditions \eqref{incond} and the Klein-Gordon equation 
\eqref{kg2}, it can be shown that:
\begin{equation*}
 \begin{split}
&\frac{\di}{\di\eta}\left.\left(\omega_k(\eta)K_k(\eta)-\sqrt{\kk^2+m^2a^2(\eta)}\right)\right|_{\eta=0}=0
\qquad\frac{\di^2}{\di\eta^2}\left.\left(\omega_k(\eta)K_k(\eta)-\sqrt{\kk^2+m^2a^2(\eta)}\right)\right|_{\eta=0}=
\frac{\left(2\kk^2+3m^2\right)^2}{\left(\kk^2+m^2\right)^\frac{3}{2}}a'^2(0)\\
&\frac{\di}{\di\eta}\left.\left(\omega_k(\eta)\Gamma_k(\eta)-\frac{\kk^2}{3\sqrt{\kk^2+m^2a^2(\eta)}}\right)\right|_{\eta=0}
=-\frac{\left(2\kk^2+3m^2\right)^2}{3\left(\kk^2+m^2\right)^\frac{3}{2}}a'(0).
 \end{split}   
\end{equation*}
Therefore, we conclude that, for a small conformal time $\Delta\eta$ after the freeze-out:
\begin{equation}\label{smalltimes2}
\begin{split}
    \left(\omega_k(\eta)K_k(\eta)-\sqrt{\kk^2+m^2a^2(\eta)}\right)&\simeq
    \frac{1}{2}\frac{\left(2\kk^2+3m^2\right)^2}{\left(\kk^2+m^2\right)^\frac{3}{2}}a'^2(0)\left(\Delta\eta\right)^2,\\
    \left(\omega_k(\eta)\Gamma_k(\eta)-\frac{\kk^2}{3\sqrt{\kk^2+m^2a^2(\eta)}}\right)&\simeq
    -\frac{\left(2\kk^2+3m^2\right)^2}{3\left(\kk^2+m^2\right)^\frac{3}{2}}a'(0)\Delta\eta,
\end{split}
\end{equation}
that is, just like for the conformal coupling, the expansion of the universe has an opposite effect 
on the quantum correction of energy density and pressure right after the freeze-out:
\begin{equation*}
    \Delta\varepsilon>0\qquad\Delta p<0.
\end{equation*}
The main difference with the conformal coupling is that, in the limit $m \to 0$, the expressions 
\eqref{smalltimes2} are still non-vanishing, so there are non-trivial corrections to the
energy density and pressure even in the massless case.

\section{Discussion and conclusions}
\label{sect6}

One may wonder what is the origin of the corrections to the classical free-streaming relations
\eqref{cfs}. In the conformal 
and minimal case, the equations \eqref{smalltimes1} and \eqref{smalltimes2} respectively, tell us 
that the corrections depend on the expansion rate $a'$ of the space-time. In fact, they appear 
not to be a pure quantum correction in that they survive the $\hbar \to 0$ limit.
In the conformal case it is clear that the deviation of $\omega_k K_k$ from $\omega_k$ is owing to the 
equations of motion of the scalar field in the FLRW spacetime. While $\omega_k K_k(0)=\omega_k(0)$, 
the derivative of $\omega_k K_k$ differs from the derivative of $\omega_k$:
$$
\frac{\di}{\di \eta} \omega_k K_k = \frac{\di}{\di \eta} \left( |v'_k|^2+\omega^2_k |v_k|^2 \right)
 = \frac{\di \omega^2}{\di \eta} |v_k|^2 \ne \frac{\di \omega}{\di \eta}
$$
and the equality applies only if $|v_k|^2 = 1/2 \omega_k$, that is in the adiabatic case \cite{mukhanov}, 
which is known to be an exact solution only if $a$ was constant. 
Now, if $\omega_k(\eta) K_k(\eta)$ evolves in time, the 
instantaneous Hamiltonian $\widehat H(\eta)$ at some time $\eta > 0$ will be of the general form 
\eqref{hami3}, ceasing to have the diagonal form \eqref{hami4} it had at the freeze-out time $\eta=0$. 
Therefore, the LLS of the Hamiltonian at time $\eta$ will no longer be the vacuum 
of the operators $\wa{({\bf k})}$, but the vacuum of a new set of operators $\wc{({\bf k})}$ which 
can be obtained by diagonalizing the \eqref{hami3} by means of suitable Bogoliubov transformations 
\eqref{bogo} (this is worked out in Appendix \ref{a3}).
The expectation value of the density of excitations with respect to the instantaneous vacuum at
the time $\eta$ can be obtained by using the equation \eqref{quadrcc} in the Appendix \ref{a3}
and reads:
\be\label{ccdev}
\begin{split}
\langle \wcd{(\mathbf{k})}\wc{(\mathbf{k})}\rangle &= |A_k|^2 \langle \wad{(\mathbf{k})}\wa{(\mathbf{k})}\rangle 
+ |B_k|^2 \langle \wa{(-\mathbf{k})}\wad{(-\mathbf{k})}\rangle \\
&=  (|A_k|^2 + |B_k|^2) n_B \left( \frac{\omega_\xi(0,k)}{T(0)} \right) + |B_k|^2 \delta^3(0) \\
&= \frac{\omega_k(\eta)}{\omega_\xi(\eta,k)} K_k(\eta) n_B \left( \frac{\omega_\xi(0,k)}{T(0)} \right) + 
 \frac{(\omega_k(\eta)/\omega_\xi(\eta,k)) K_k(\eta) -1}{2} \delta^3(0) 
\end{split}
\ee
where we have used eqs.~\eqref{valasp},\eqref{hyp},\eqref{bogosol} and we have taken advantage 
of the invariance by reflection ${\bf k} \to -{\bf k}$ of the Bose-Einstein distribution \eqref{bose}. 
Therefore, by comparing the expression of the energy density \eqref{EeP} with the equation 
\eqref{ccdev}, one would be led to interpret the enhancement of the energy density with the continuous 
creation of excitations with respect to the vacuum at the time $\eta$ 
\cite{ford,nippon,redi} because the instantaneous LLS of the Hamiltonian $\widehat H$ changes with time.
Nevertheless, the pressure in the equation \eqref{EeP} is not simply the classical expression of
pressure \eqref{cfs} multiplied by the same factor $K_k(\eta)$.

Albeit the corrections \eqref{quantumcor} and \eqref{quantumcor2} are not proportional to 
$\hbar$, they do ensue from the quantization of the Klein-Gordon field and the use of the quantum 
statistical operator \eqref{densop} expressed in terms of quantum fields.

In conclusion, we have determined the equation of state of neutral scalar freely streaming matter 
after decoupling by using the equation of motion of the field in curved space-time within a quantum 
statistical mechanics framework. By using the density operator of local thermodynamic equilibrium at 
the decoupling, we have obtained some peculiar corrections in the equation of state with respect to 
the predictions of classical relativistic kinetic theory. 
Right after the freeze-out, the quantum field correction to the energy density is positive and can
be possibly interpreted as a particle creation term due to cosmological expansion, whereas the quantum field
correction to pressure is negative. 
In the long run, the quantitative relevance of such corrections with respect to the energy density 
and pressure as a whole, depends on the specific solution of the field equation, hence on the entire 
evolution of the scale factor $a(t)$. As has been discussed, this is impossible to assess 
without solving the Klein-Gordon equation, which in turn requires the knowledge of the function $a(\eta)$,
and a dedicated study is in preparation \cite{becarose2}. If the quantum corrections 
were comparable to the classical terms at long times, there would likely be consequences also for the 
statistical fluctuations of energy density and pressure, hence for the problem of large scale structure
formation.

\section*{Acknowledgments}

We acknowledge interesting discussions with D. Glavan, M. Redi, D. Seminara



\appendix
 
\section{The stress-energy tensor of free the scalar field}\label{A1}

We start rewriting the stress-energy tensor operator  \eqref{setkg} of the scalar field as:
\begin{equation}\label{tmunu}
\begin{split}
    \widehat{T}_{\mu\nu}=&\nabla_\mu\widehat{\psi}\nabla_\nu\widehat{\psi}
    -\frac{1}{2}g_{\mu\nu}\nabla_\rho\widehat{\psi}\nabla^\rho\widehat{\psi}+
    \frac{1}{2}m^2\widehat{\psi}^2 \\
    &+2\xi\left(-\nabla_\mu\widehat{\psi}\nabla_\nu\widehat{\psi}+g_{\mu\nu}
    \nabla_\rho\widehat{\psi}\nabla^\rho\widehat{\psi}+g_{\mu\nu}\widehat{\psi}\Box\widehat{\psi}
    -\widehat{\psi}\nabla_\mu\nabla_\nu\widehat{\psi}+\frac{G_{\mu\nu}}{2}\widehat{\psi}^2\right).
\end{split}
\end{equation}
whose time-time component in conformal coordinates reads:
\begin{equation}\label{PSItzero}
        \widehat{T}_{00}=\frac{1}{2}\widehat{\psi}'^2+\frac{1}{2}\vec{\nabla}\widehat{\psi}
        \cdot\vec{\nabla}\widehat{\psi}+\frac{1}{2}m^2a^2\widehat{\psi}^2
        +2\xi\left[-\vec{\nabla}\widehat{\psi}\cdot\vec{\nabla}\widehat{\psi}+a^2\xi R\widehat{\psi}^2-
        m^2a^2\widehat{\psi}^2-\widehat{\psi}\widehat{\psi}''+
        \frac{a'}{a}\widehat{\psi}\widehat{\psi}'+\frac{G_{00}}{2}\widehat{\psi}^2\right]
\end{equation}
In the equation \eqref{PSItzero} we have used the Klein-Gordon equation \eqref{kg}. 
By using the rescaled field $\widehat{\chi}=a\widehat{\psi}$, with the relations:
\begin{equation*}
    \begin{split}
        \widehat{\psi}=\frac{\widehat{\chi}}{a},\quad
        \widehat{\psi}'=\frac{\widehat{\chi}'}{a}-\widehat{\chi}\frac{a'}{a^2},\quad
        \widehat{\psi}''=\frac{\widehat{\chi}''}{a}-2\widehat{\chi}'\frac{a'}{a^2}-
        \widehat{\chi}\left( \frac{a''}{a^2}-\frac{2a'^2}{a^3} \right).
    \end{split}
\end{equation*}
the \eqref{PSItzero} becomes:
\begin{equation}\label{mezzotensore}
\begin{split}
    \widehat{T}_{00}&=\frac{1}{2a^2}\left[\widehat{\chi}'^2+\vec{\nabla}\widehat{\chi}\cdot
    \vec{\nabla}\widehat{\chi}+\widehat{\chi}^2\left(m^2a^2+\frac{a'^2}{a^2}\right)
    -2\widehat{\chi}\widehat{\chi}'\frac{a'}{a}\right]+\\
    &+\frac{2\xi}{a^2}\left[-\vec{\nabla}\widehat{\chi}\cdot\vec{\nabla}\widehat{\chi}+
    a^2\xi R\widehat{\chi}^2-m^2a^2\widehat{\chi}^2+\frac{G_{00}}{2}\widehat{\chi}^2+3\widehat{\chi}
    \widehat{\chi}'\frac{a'}{a}-3\widehat{\chi}^2\frac{a'^2}{a^2}+\widehat{\chi}^2\frac{a''}{a}
    -\widehat{\chi}\widehat{\chi}''\right].
\end{split}
\end{equation}
and the Klein-Gordon equation now reads \cite{mukhanov}:
\begin{equation}
    \widehat{\chi}''-\vec{\nabla}^2\widehat{\chi}+\left[m^2a^2-(1-6\xi)\frac{a''}{a}\right]\widehat{\chi}=0,
\end{equation}
where $\vec{\nabla}^2=\left(\frac{\partial^2}{\partial x^2},\frac{\partial^2}{\partial y^2},
\frac{\partial^2}{\partial z^2}\right)$ is the Laplacian operator.

In conformal coordinates the time-time components of the Einstein tensor and the curvature
scalar are:
\begin{equation}\label{RicciEinstein}
    G_{00}=\frac{3a'^2}{a^2},\ R=-\frac{6a''}{a^3}.
\end{equation}
Rewriting the second derivative of the field in terms of the Laplacian and plugging the expressions 
\eqref{RicciEinstein} into the \eqref{mezzotensore} we obtain the final expression for the time-time
component of the energy-momentum tensor \eqref{tmunu} in the conformal coordinates in terms of 
the $\chi$ field:
\begin{equation}\label{T00APP}
    \widehat{T}_{00}=\frac{1}{2a^2} \left[ \widehat{\chi}'^2+\vec{\nabla}\widehat{\chi}\cdot\vec{\nabla}
    \widehat{\chi}+\widehat{\chi}^2\left( m^2a^2+(1-6\xi)\frac{a'^2}{a^2}\right)
    -(1-6\xi)2\widehat{\chi}'\widehat{\chi}\frac{a'}{a}-4\xi \left(\widehat{\chi}\vec{\nabla}^2\widehat{\chi}
    +\vec{\nabla}\widehat{\chi}\cdot\vec{\nabla}\widehat{\chi}\right)\right].
\end{equation}
To obtain the same component in Cartesian coordinates, it is sufficient to divide by $a^2$ and
thereby the equation \eqref{tzero} is recovered.

The other ingredient which is needed to calculate the pressure is the spacial trace of the stress-energy
tensor, that is, in conformal coordinates:
\be\label{strace}
    \sum_{i=1}^3 \widehat{T}_{ii}= \widehat{T}_{00}-a^2 \widehat{T}^{\mu}_\mu,
\ee
where $\wT^\mu_\mu$ is the trace of the stress-energy tensor which turns out to be, by using 
the eq.~\eqref{tmunu}:
$$
    \widehat{T}^\mu_\mu=\left( 3\xi-\frac{1}{2} \right)\Box\widehat{\psi}^2+m^2\widehat{\psi}^2+
    \widehat{\psi}\left( \Box+m^2-\xi R \right)\widehat{\psi}.
$$
Using again the equations of motion \eqref{kg} and reintroducing the $\widehat{\chi}$ field, the
trace becomes:
\begin{equation}\label{trace}
    \widehat{T}^\mu_\mu=\frac{1}{a^4}\left[ 2m^2 a^2 \widehat{\chi}^2 (1-3\xi) + (6 \xi -1)
    \left( \widehat{\chi}'^2 - \vec{\nabla}\widehat{\chi}\cdot\vec{\nabla}\widehat{\chi}
    -2 \widehat{\chi}'\widehat{\chi}\frac{a'}{a} + \wchi^2 \frac{a'^2}{a^2} +\xi R a^2 \wchi^2 
    \right) \right]
\end{equation}
Finally, the spatial trace \eqref{strace}, by using the \eqref{T00APP} and the \eqref{trace}, 
turns out to be:
\begin{equation*}
        \sum_{i=1}^3\widehat{T}_{ii}=\frac{1}{2a^2}\left[3\left(1-4\xi\right)
        \left(\widehat{\chi}'^2-m^2a^2\widehat{\chi}^2\right)+\left(12\xi-1\right)
        \vec{\nabla}\widehat{\chi}\cdot\vec{\nabla}\widehat{\chi}-4\xi\left(\widehat{\chi}
        \vec{\nabla}^2\widehat{\chi}+\vec{\nabla}\widehat{\chi}\cdot\vec{\nabla}\widehat{\chi}\right)\\
        +3\left(1-6\xi\right)\left(\frac{a'^2}{a^2}-4\xi\frac{a''}{a}\right)\right],
\end{equation*}
which coincides with \eqref{tii}.

\section{Calculation of the Hamiltonian}\label{A2}

In this Section we demonstrate that the integration of the operator \eqref{tzero} over the space-like 
hypersurface at fixed cosmological time yields the Hamiltonian \eqref{hami3}. To do this, 
we integrate over $\di^3\mathbf{x}$ each quadratic combination of the field operators appearing
in the equation \eqref{tzero}. We start with:
\begin{equation}\label{chi2}
\begin{split}
    \int \di^3 \x \;\widehat{\chi}^2&=\frac{1}{(2\pi)^3}\int \di^3 \x \int \di^3 \kk
    \int \di^3 \kk \;'\left[ \left( \e^{\ii \mathbf{x}\cdot\mathbf{k}} v_k \, \widehat{a}(\mathbf{k})
    +\e^{-\ii\mathbf{k}\cdot\mathbf{x}}v^*_k \, \widehat{a}^\dagger(\mathbf{k}) \right)
    \left( \e^{\ii\mathbf{x}\cdot\mathbf{k}'} v_{k'} \, \widehat{a}(\mathbf{k}')+
    \e^{-\ii\mathbf{k}'\cdot\mathbf{x}} v^*_{k'} \, \widehat{a}^\dagger(\mathbf{k}') \right)\right]\\
    &=\int \di^3 \kk \;\left[v^2_k \, \widehat{a}(\mathbf{k})\widehat{a}(-\mathbf{k})+
    (v^*_k)^2 \, \widehat{a}^\dagger(\mathbf{k})\widehat{a}^\dagger(-\mathbf{k})+|v_k|^2 \,
    \left( \widehat{a}^\dagger(\mathbf{k}) \widehat{a}(\mathbf{k})+ \widehat{a}(\mathbf{k})
     \widehat{a}^\dagger (\mathbf{k}) \right)\right],
\end{split}
\end{equation}
where we have used:
\begin{equation*}
    \int \di^3 \x \; \e^{\ii\mathbf{x}\cdot(\mathbf{k}\pm\mathbf{k}')}=
    (2\pi)^3\delta^3(\mathbf{k}\pm\mathbf{k}'),
\end{equation*}
The second quadratic term in the derivative with respect to the conformal time of the $\wchi$ field, 
can be obtained by just replacing the functions $v_k$ with their derivatives in the \eqref{chi2}: 
\begin{equation}\label{chipr2}
    \int \di^3 \x \; \widehat{\chi}'^2=
    \int \di^3 \kk \;\left[(v^{\prime}_k)^2 \, \widehat{a}(\mathbf{k})\widehat{a}(-\mathbf{k})+
    (v^{\prime *}_k)^2 \, \widehat{a}^\dagger(\mathbf{k})\widehat{a}^\dagger(-\mathbf{k})+|v'_k|^2\,
    \left(\widehat{a}^\dagger(\mathbf{k})\widehat{a}(\mathbf{k})+\widehat{a}(\mathbf{k})
    \widehat{a}^\dagger(\mathbf{k})\right)\right].
\end{equation}
Similarly, one can obtain the mixed term:
\begin{equation}\label{chiprchi}
    \int \di^3\x \; \widehat{\chi}\widehat{\chi}'= \int \di^3 \kk \; 
    \left[\widehat{a}(\mathbf{k})\widehat{a}(-\mathbf{k})v'_kv_k+\widehat{a}^\dagger(\mathbf{k})
    \widehat{a}^\dagger(-\mathbf{k})v^{\prime *}_kv^*_k
    +\left(\widehat{a}^\dagger(\mathbf{k})\widehat{a}(\mathbf{k})+
    \widehat{a}(\mathbf{k})\widehat{a}^\dagger(\mathbf{k})\right)(v'^*_kv_k+v'_kv^*_k)\right].
\end{equation}

We now turn to terms containing the spatial derivatives of the field. We have:
\begin{equation*}
    \Vec{\nabla}\widehat{\chi}=\frac{1}{(2\pi)^{\frac{3}{2}}}\int \di^3 \kk \; 
    \ii \mathbf{k}\left[\e^{i\mathbf{k}\cdot\mathbf{x}}v_k \, \widehat{a}(\mathbf{k})-
    \e^{-\ii \mathbf{k}\cdot\mathbf{x}} v^*_k \, \widehat{a}^\dagger(\mathbf{k})\right],
\end{equation*}
whence:
\begin{equation}\label{grad2}
\begin{split}
\int \di^3\x\left|\Vec{\nabla}\widehat{\chi}\right|^2 &= \frac{1}{(2\pi)^3} \int \di^3\x \int \di^3 \kk 
\int \di^3 \kk' \; \left[\ii \mathbf{k}\left( \e^{\ii \mathbf{k}\cdot\mathbf{x}}v_k \, \widehat{a}(\mathbf{k})
-\e^{-\ii \mathbf{k}\cdot\mathbf{x}}v^*_k\, \widehat{a}^\dagger(\mathbf{k}) \right) (-\ii \mathbf{k}') 
\left( \e^{-i\mathbf{k}\cdot\mathbf{x}} v^*_{k'} \, \widehat{a}^\dagger(\mathbf{k}')-
\e^{\ii \mathbf{k}\cdot\mathbf{x}} v_k \, \widehat{a}(\mathbf{k}') \right)\right]\\
&=\int \di^3 \kk \; \kk^2 \left[ (v_k)^2 \, \widehat{a}(\mathbf{k})\widehat{a}(-\mathbf{k})+
(v^*_k)^2\, \widehat{a}^\dagger(\mathbf{k})\widehat{a}^\dagger({-\mathbf{k}})+ |v_k|^2 \, 
\left(\widehat{a}^\dagger(\mathbf{k})\widehat{a}(\mathbf{k})+\widehat{a}(\mathbf{k}) 
\widehat{a}^\dagger(\mathbf{k})\right)\right].
\end{split}
\end{equation}
We are now concerned with the last term in \eqref{tzero} involving the Laplacian. Since:
\begin{equation*}
    \widehat{\chi}\vec{\nabla}^2\widehat{\chi}+\vec{\nabla}\widehat{\chi}\cdot\vec{\nabla}\widehat{\chi}
    = \vec{\nabla} \cdot \left( \widehat{\chi} \vec{\nabla} \widehat{\chi} \right)
\end{equation*}
the integration yields a surface term which vanishes. This can be checked by plugging the field 
expansion \eqref{field} and verifying that each mode generates two terms with opposite sign.
 
In conclusion, by adding the integrals \eqref{chi2}, \eqref{chipr2}, \eqref{chiprchi} and \eqref{grad2}
with the suitable coefficients read off from the equation \eqref{tzero}, we obtain the expression for 
the effective Hamiltonian operator defined by the equation \eqref{hami} (using the Cartesian coordinates):
\begin{equation}
\begin{split}
\widehat{H} &= a^3 \int \di^3 \x \; \wT^{00}(x) = a^3 \int \di^3 \x \; \wT_{00}(x) \\
&= \frac{1}{2a}\int \di^3 \kk \; \left\{\widehat{a}(\mathbf{k})\widehat{a}(-\mathbf{k})\left[v'^2_k
+ (\kk^2+m^2a^2)v^2_k+(1-6\xi)\left( \frac{a'^2}{a^2}v^2_k-\frac{a'}{a}v'_kv_k \right)\right]\right.\\
&+\widehat{a}^\dagger(\mathbf{k})\widehat{a}^\dagger(-\mathbf{k})\left[(v'^*_k)^2+(\kk^2+m^2a^2)(v^*_k)^2+
(1-6\xi)\left( \frac{a'^2}{a^2}v^2_k-\frac{\alpha'}{\alpha}v'^*_kv^*_k \right)\right]\\
&+\widehat{a}^\dagger(\mathbf{k})\widehat{a}(\mathbf{k})\left[\left|v'_k\right|^2+(\kk^2+m^2a)\left|v_k\right|^2
+(1-6\xi)\left( \frac{a'^2}{a^2}\left|v_k\right|^2-\frac{a'}{a}(v'^*_kv_k+v'_kv^*_k) \right)\right]\\
&\left.+\widehat{a}(\mathbf{k})\widehat{a}^\dagger(\mathbf{k})\left[\left|v'_k\right|^2+(\kk^2+m^2a)
\left|v_k\right|^2+(1-6\xi)\left( \frac{a'^2}{a^2}\left|v_k\right|^2-\frac{a'}{a}(v'^*_kv_k+v'_kv^*_k)
\right)\right]\right\},
\end{split}
\end{equation}
which, introducing the functions \eqref{funz}, reduces to:
\begin{equation}
    \widehat{H}(\eta)=\frac{1}{2a(\eta)}\int \di^3 \kk \;\omega_k(\eta)\left[K_k(\eta)
    \left(\widehat{a}^\dagger(\mathbf{k})\widehat{a}(\mathbf{k})+
    \widehat{a}(\mathbf{k})\widehat{a}^\dagger(\mathbf{k})\right)+\Lambda_k(\eta)
    \widehat{a}(\mathbf{k})\widehat{a}(-\mathbf{k})+\Lambda^*_k(\eta)\widehat{a}^\dagger(\mathbf{k})
    \widehat{a}^\dagger(-\mathbf{k})\right],
\end{equation}
coinciding with the eq.~\eqref{hami3}.

\section{Covariant conservation of the stress-energy tensor}\label{a4}

Given the perfect fluid form of the stress-energy tensor (\ref{set}) the covariant conservation 
in the FLRW metric implies, as it is well known:
\begin{equation}\label{ConservationEq}
   \varepsilon'=-\frac{3a'}{a}\left(\varepsilon+p\right).
\end{equation}
where the prime is the derivative with respect the conformal time $\eta$.
The energy density and the pressure (\ref{EeP}) can be written as:
\begin{equation*}
    \begin{split}
      \varepsilon(\eta)=\int\di^3\kk \; \varepsilon_k(\eta),\qquad p(\eta)=\int\di^3\kk \; p_k(\eta),
    \end{split}
\end{equation*}
where $\varepsilon_k$ and $p_k$ are the integrands of the of the (\ref{EeP}):
\begin{equation}\label{IntegrandEpsiloneP}
    \varepsilon_k(\eta)\equiv\frac{n_B(0)}{\left(2\pi\right)^3}\frac{\omega_k(\eta)K_k(\eta)}
    {a^4(\eta)},\quad p_k(\eta)\equiv\frac{n_B(0)}{\left(2\pi\right)^3}
    \frac{\omega_k(\eta)\Gamma_k(\eta)}{a^4(\eta)}.
\end{equation}
The conservation equation (\ref{ConservationEq}) can be indeed checked for each $k$. Plugging 
the (\ref{IntegrandEpsiloneP}) in the (\ref{ConservationEq}) one obtains the relation:
\begin{equation}\label{RelazioneDaDimostrare}
    \frac{\di}{\di\eta}\left[\omega_k(\eta)K_k(\eta)\right]=\frac{a'(\eta)}{a(\eta)}
    \left[\omega_k(\eta)K_k(\eta)-3\omega_k(\eta)\Gamma_k(\eta)\right].
\end{equation}
We are going to show that the above equation is fulfilled, so that the covariant conservation 
\eqref{ConservationEq} holds. Let us start by calculating the left hand side of the 
\eqref{RelazioneDaDimostrare}:
\begin{equation*}
    \begin{split}
        \frac{\di}{\di\eta}\left(\omega_kK_k\right)&=2\left(1-6\xi\right)\frac{a'^2}{a^2}
        \mbox{Re}\left(v'_kv^*_k\right)+2\left|v_k\right|^2\left[\omega_k\omega'_k+\left(1-6\xi\right)
        \frac{a'^2}{a^2}\Omega^2_k\right]-2\left(1-6\xi\right)\frac{a'}{a}\left|v'_k\right|^2,
    \end{split}
\end{equation*}
where we used the expressions \eqref{funz} and the equation of motion \eqref{kg2}. Since:
\begin{equation*}
    \omega'_k=\frac{1}{\omega_k}\left[m^2a'a+\left(1-6\xi\right)\frac{a'^2}{a^2}
    \left(\frac{a''}{a'}-\frac{a'}{a}\right)\right],
\end{equation*}
and using the expression \eqref{kg2} for $\Omega^2_k$, we get:
\begin{equation}\label{lhscurly}
    \begin{split}
        \frac{\di}{\di\eta}\left(\omega_kK_k\right)=\frac{a'}{a}&\left\{2\left(1-6\xi\right)
        \left[\frac{a'}{a}\mbox{Re}\left(v'_kv^*_k\right)-\left|v'_k\right|^2\right]\right.\\
        &\left.+2\left|v_k\right|^2\left[\left(1-6\xi\right)\left(\kk^2+6\xi\frac{a''}{a}
        -\frac{a'^2}{a^2}\right)+2m^2a^2\left(1-3\xi\right)\right]\right\}.
    \end{split}
\end{equation}
We now turn to the right hand side of the \eqref{RelazioneDaDimostrare}. We have:
\begin{equation*}
\begin{split}
    \omega_kK_k-3\omega_k\Gamma_k&=\left|v'_k\right|^2\left(-2+12\xi\right)+
    \left|v_k\right|^2\left(\omega^2_k-\gamma_k\right)+2\left(1-6\xi\right)\mbox{Re}\left(v'_kv^*_k\right).
\end{split}
\end{equation*}
Using the expression (\ref{gamma}) for $\gamma_k$:
\begin{equation*}
\begin{split}
     \omega^2_k-\gamma_k&=2\left[\left(1-6\xi\right)\left(\kk^2-\frac{a'^2}{a^2}
     +6\xi\frac{a''}{a}\right)+2m^2a^2\left(1-3\xi\right)\right],
\end{split}
\end{equation*}
so that:
\begin{equation*}
\begin{split}
    \omega_kK_k-3\omega_k\Gamma_k&=2\left(1-6\xi\right)\left[\frac{a'}{a}\mbox{Re}
    \left(v'_kv^*_k\right)-\left|v'_k\right|^2\right]\\
        &+2\left|v_k\right|^2\left[\left(1-6\xi\right)\left(\kk^2+6\xi\frac{a''}{a}-
        \frac{a'^2}{a^2}\right)+2m^2a^2\left(1-3\xi\right)\right],
\end{split}
\end{equation*}
The right hand side of the above equation is precisely the expression within the curly brackets 
in the equation \eqref{lhscurly}. Therefore, the equation \eqref{RelazioneDaDimostrare}) holds 
and the \eqref{ConservationEq} as a consequence thereof. 

\section{Bogoliubov relations and diagonalization of the Hamiltonian}\label{a3}

Our goal is to make the Hamiltonian \eqref{hami3} of the diagonal  form \eqref{hami4} with a suitable 
choice of the creation and annihilation operators $\wc{({\bf k})},\wcd{({\bf k})}$
related to the $\wa{({\bf k})}$ and $\wad{({\bf k})}$ by the Bogoliubov relations \eqref{bogo}. 

The equation \eqref{klcons}, and the reality of $K_k(\eta)$ makes it possible to set:
\begin{equation}\label{hyp}
\begin{split}
K_k(\eta)&=\frac{\omega_\xi(\eta,k)}{\omega_k(\eta)}\cosh\left[2\Theta_k(\eta)\right],\\
\Lambda_k(\eta)&=\frac{\omega_\xi(\eta,k)}{\omega_k(\eta)}\sinh\left[2\Theta_k(\eta)\right]
\exp\left[i\Phi_k(\eta)\right].
\end{split}
\end{equation}
which is a very useful parametrization of the two functions $K_k(\eta)$ and $\Lambda_k(\eta)$ in
terms of an hyperbolic angle $\Theta_k(\eta)$ and a phase $\Phi_k(\eta)$. A new set of creation
and annihilation operators can be defined in terms of the so-called Bogoliubov transformations 
(see eq.~\eqref{bogo}):
\be\label{bogo1}
\begin{split}
\wc{(\mathbf{k})}& = A^*_k \wa{(\mathbf{k})} - B^*_k \wad{(-\mathbf{k})},\\
\wcd{(\mathbf{k})}& = A_k \wad{(\mathbf{k})}  - B_k \wa{(-\mathbf{k})},
\end{split}
\ee
They fulfill standard commutation relations for, as it was pointed out in the Section~\ref{sect3}, 
the coefficients $A_k$ and $B_k$ fulfill the relations:
\be\label{constraint}
    \left| A_k \right|^2 - \left| B_k \right|^2=1,
\ee
By using the eq.~\eqref{constraint}, the Bogoliubov relations \eqref{bogo1} can be easily inverted:
\be\label{bogo2}
\begin{split}
\wa{(\mathbf{k})}& = A_k \wc{({\bf k})} + B^*_k \wcd{(-\mathbf{k})},\\
\wad{(\mathbf{k})}& = A^*_k \wcd{(\mathbf{k})} + B_k \wc{(-\mathbf{k})},
\end{split}
\end{equation}

We can rewrite the Hamiltonian \eqref{hami3} in terms of then new set of creation and annhilation operators $\left\{\widehat{c}(\mathbf{k}),\widehat{c}^\dagger(\mathbf{k})\right\}$ by using the relations \eqref{bogo2} 
and changing, when needed, the integration variable from $-{\bf k}$ to ${\bf k}$ keeping in mind that 
$\omega_k,\Lambda_k,K_k$ are all even functions of ${\bf k}$:
\begin{equation}\label{newhami}
\begin{split}
    \widehat{H}(\eta) &=\frac{1}{2a(\eta)}\int \di^3 \kk \; 
    \omega_k(\eta) \left[ \left( \wcd{(\mathbf{k})}\wc{(\mathbf{k})} + \wc{(-\mathbf{k})}\wcd{(-\mathbf{k})} \right)
    \left( K_k(\eta) |A_k|^2 + K_k (\eta) |B|^2 + A_k^* B_k \Lambda_k(\eta)^* + A_k B_k^* \Lambda_k(\eta) \right)\right. \\
  &+ \left.\wcd{(\mathbf{k})}\wcd{-(\mathbf{k})} \left( 2 K_k(\eta) A_k^* B_k^*+ \Lambda_k(\eta)^* A_k^{*2} 
  + \Lambda_k B_k^{*2} \right) + \wc{(\mathbf{k})}\wc{-(\mathbf{k})} \left( 2 K_k(\eta) A_k B_k + 
  \Lambda_k(\eta) A_k^{2} + \Lambda_k^* B_k^{2} \right) \right]
\end{split}
\end{equation}
In order to diagonalize the operator \eqref{newhami} we must then impose the following condition, for
each $k$:
\be\label{diageq}
    2 K_k(\eta) A_k B_k + \Lambda_k(\eta) A_k^2 + \Lambda_k(\eta)^* B_k^{2} = 0.
\ee
We can set, in view of the constraint \eqref{constraint}:
$$
    A_k =\cosh \theta_k \e^{\ii  \varphi_A} \qquad \qquad B_k = \sinh \theta_k \, \e^{\ii \varphi_B}.
$$
Using the hyperbolic parametrization for $A_k,\ B_k$ and for $K_k,\ \Lambda_k$, the \eqref{diageq}
becomes:
\be\label{diageq2}
    2 \cosh 2\Theta \cosh \theta \sinh \theta \, \e^{\ii(\varphi_A+\varphi_B)} 
    + \sinh 2\Theta \, \e^{\ii\Phi} \cosh^2 \theta \, \e^{2 \ii \varphi_A} + 
    \sinh 2\Theta \sinh^2\theta \, \e^{-\ii \Phi}  \e^{2\ii\varphi_B} = 0
\ee
This is one complex equation for three real unknowns. Indeed, one global phase factor remains 
undetermined in the Bogoliubov transformation because a global phase factor can be always reabsorbed
in a trivial redefinition of the free particle states. Hence, we can then assume $A_k$ to 
be real by setting $\varphi_A = 0$. With this position, it can be readily verified that the solution 
of the \eqref{diageq2} is:
$$
    \theta= - \Theta \qquad \qquad  \varphi_B=\Phi,
$$
We can then write the solution for the Bogoliubov coefficients in the eqs.~\eqref{bogo1},\eqref{bogo2}:
\be\label{bogosol}
A_k(\eta) = \cosh \Theta_k(\eta) \qquad \qquad B_k(\eta) = - \sinh \Theta_k(\eta) \e^{\ii \Phi_k(\eta)}
\ee
with $\Theta_k(\eta)$ and $\Phi_k(\eta)$ defined by the equation \eqref{hyp}.
With this solution, the coefficient of the combination 
$\wcd{(\mathbf{k})}\wc{(\mathbf{k})} + \wc{(-\mathbf{k})}\wcd{(-\mathbf{k})}$ in the equation~\eqref{newhami} 
becomes, by using the \eqref{hyp} and the \eqref{bogosol}:
\begin{align*}
 & K_k(\eta)(|A_k|^2+|B_k|^2) + 2 \mbox{Re}(\Lambda_k(\eta)A_k B^*_k) = \\
 & = \frac{\omega_\xi(\eta,k)}{\omega_k(\eta)} \cosh^2 2\Theta_k(\eta) - 2 \frac{\omega_\xi(\eta,k)}{\omega_k(\eta)}
  \sinh 2\Theta_k (\eta) \sinh \Theta_k(\eta) \cosh\Theta_k(\eta) = \frac{\omega_\xi(\eta,k)}{\omega_k(\eta)}
\end{align*}
and so the operator \eqref{hami3} becomes:
\begin{equation}\label{Hdiag}
    \widehat{H}(\eta)=\frac{1}{a(\eta)}\int \di^3 \kk \;\omega_\xi(\eta,k)
    \left( \wcd{(\mathbf{k})}\wc{(\mathbf{k})}+\frac{1}{2} \right).
\end{equation}
It is also interesting to find the quadratic combination of creation and annihilation operator in
the new set. By using the eq.~\eqref{bogo1} and the eq.~\eqref{bogosol}, we have:
\be\label{quadrcc}
\begin{split}
\wcd{(\mathbf{k})}\wc{(\mathbf{k})} &= |A_k|^2 \wad{(\mathbf{k})}\wa{(\mathbf{k})} +
|B_k|^2 \wa{(-\mathbf{k})}\wad{(-\mathbf{k})} - A_k B^*_k \wad{(\mathbf{k})}\wad{(-\mathbf{k})}
- A^*_k B_k \wa{(-\mathbf{k})}\wa{(\mathbf{k})}  \\
& = \cosh^2 \Theta_k(\eta) \wad{(\mathbf{k})}\wa{(\mathbf{k})} + \sinh^2 \Theta_k(\eta)
 \wa{(-\mathbf{k})}\wad{(-\mathbf{k})} \\
 & + \cosh \Theta_k(\eta) \sinh \Theta_k(\eta)\, \e^{-\ii \Phi_k(\eta)}
 \wad{(\mathbf{k})}\wad{(-\mathbf{k})} + \cosh \Theta_k(\eta) \sinh \Theta_k(\eta)\, \e^{\ii \Phi_k(\eta)}
 \wa{(-\mathbf{k})}\wa{(\mathbf{k})} 
\end{split}
\ee
which can be used to determine the density of excitations with respect to the vacuum which
is annihilated by the new annihilation operators $\wc{({\bf k})}$.


\begin{thebibliography}{99}
\section*{References}

\bibitem{dowker1}
J.~S.~Dowker and G.~Kennedy,
J. Phys. A \textbf{11}, 895 (1978).

\bibitem{dowker2}
J.~S.~Dowker and J.~P.~Schofield,
Phys. Rev. D \textbf{38}, 3327 (1988).

\bibitem{critchley}
B.~L.~Hu, R.~Critchley and A.~Stylianopoulos,
Phys. Rev. D \textbf{35}, 510 (1987).

\bibitem{miele}
D.~V.~Fursaev and G.~Miele,
Phys. Rev. D \textbf{49}, 987-998 (1994).

\bibitem{kolbturner}
E.~W.~Kolb and M.~S.~Turner, Front. Phys. \textbf{69}, 1-547 (1990).

\bibitem{kremer}
C. Cercignani, G.~M.~Kremer, {\it The Relativistic Boltzmann Equation: Theory and Applications},
Prog. Math. Phys. {\bf 22}, Birkh\"auser Verlag, 2002.

\bibitem{piattella}
O.~F.~Piattella, {\it Lecture Notes in Cosmology}, Springer, 2018,
[arXiv:1803.00070 [astro-ph.CO]].

\bibitem{fonarev1}
O.~A.~Fonarev,
[arXiv:gr-qc/9311018 [gr-qc]].
  
\bibitem{fonarev2}
O.~A.~Fonarev,
Phys. Lett. A \textbf{190}, 29 (1994).

\bibitem{allen}
B.~Allen, A.~Folacci and G.~W.~Gibbons,
Phys. Lett. B \textbf{189}, 304 (1987).

\bibitem{ambrus}
V.~E.~Ambrus and E.~Winstanley,
Class. Quant. Grav. \textbf{34}, no.14, 145010 (2017).

\bibitem{becarose2}
F. Becattini, D. Roselli, in preparation.

\bibitem{birrell} N.~D.~Birrell and P.~C.~W.~Davies,
{\it Quantum Fields in Curved Space}, Cambridge Univ. Press, 1984 , and references
therein.

\bibitem{zubarev}
D.~N.~Zubarev, A.~V.~Prozorkevich, S.~A.~Smolyanskii, Theoret. and Math. 
Phys. 40, 821 (1979).
 
\bibitem{weert}
Ch.~G.~Van~Weert, Ann.\ Phys.\ {\bf 140}, 133 (1982).
 
\bibitem{betaframe}
F.~Becattini, L.~Bucciantini, E.~Grossi and L.~Tinti,  Eur.\ Phys.\ J.\ C {\bf 75}, 
no. 5, 191 (2015).

\bibitem{hosoya}
A.~Hosoya, M.~Sakagami and M.~Takao, Annals Phys.\  {\bf 154}, 229 (1984).

\bibitem{becazuba}
F.~Becattini, M.~Buzzegoli and E.~Grossi, Particles \textbf{2}, no.2, 197-207 (2019).

\bibitem{glavan}
D.~Glavan, T.~Prokopec and V.~Prymidis,
Phys. Rev. D \textbf{89}, no.2, 024024 (2014)

\bibitem{glavan2}
D.~Glavan, T.~Prokopec and T.~Takahashi,
Phys. Rev. D \textbf{94}, 084053 (2016).

\bibitem{degroot}
S.~R.~De Groot, W.~A.~Van Leeuwen and C.~G.~Van Weert,
{\it Relativistic Kinetic Theory. Principles and Applications} North Holland, 1980.

\bibitem{mukhanov}
V.~Mukhanov and S.~Winitzki, {\it Introduction to quantum effects in gravity},
Cambridge University Press, 2007.

\bibitem{ford}
L.~H.~Ford,
Phys. Rev. D \textbf{35}, 2955 (1987).

\bibitem{nippon}
Y.~Ema, K.~Nakayama and Y.~Tang, JHEP \textbf{09}, 135 (2018).

\bibitem{redi}
M.~Redi and A.~Tesi,
arXiv:2210.03108 [hep-ph].


\end{thebibliography}
\end{document}